\documentclass[12pt]{article}
\usepackage[cp866]{inputenc}
\usepackage{amssymb,cite}
\usepackage{indentfirst}

\newcommand{\be}{\begin{eqnarray}}
\newcommand{\ee}{\end{eqnarray}}
\newcommand{\lab}{\label}
\newcommand{\ep}{ $  e^+ e^- $ }
\newcommand{\n}{\nu,\tilde\nu}

\newcommand{\appendixes}{
\par\setcounter{section}{0}
\setcounter{subsection}{0}
\setcounter{equation}{0}
\def\thesection{ Appendix \Alph{section} }
\def\theequation{\Alph{section}.\arabic{equation} }
}

\inputencoding{cp866}

\begin{document}

\title{
Neutrino Interaction with Nucleons in the Envelope 
of a Collapsing Star with a Strong Magnetic Field
}

\author{A.~A.~Gvozdev{\footnote{e-mail: gvozdev@univ.uniyar.ac.ru}},
I.~S.~Ognev{\footnote{e-mail: ognev@univ.uniyar.ac.ru}} \\[2mm]
{\it Yaroslavl State University }\\[2mm]
{\it ul. Sovetskaya 14, Yaroslavl, 150000 Russia }}

\date{}

\maketitle

\begin{abstract}
{\normalsize
The interaction of neutrinos with nucleons in the envelope of a remnant of 
collapse with a strong magnetic field during the passage of the main 
neutrino flux is investigated. General expressions are derived for the 
reaction rates and for the energy-momentum transferred to the medium 
through the neutrino scattering by nucleons and in the direct URCA 
processes. Parameters of the medium in a strong magnetic field are 
calculated under the condition of quasi-equilibrium with neutrinos. 
Numerical estimates are given for the neutrino mean free paths and for the 
density of the force acting on the envelope along the magnetic field. It 
is shown that in a strong toroidal magnetic field, the envelope region 
partially transparent to neutrinos can acquire a large angular 
acceleration on the passage time scales of the main neutrino flux. 
}
\end{abstract}


\section{
Introduction
}

Collapsing systems with rapidly rotating remnants are of great interest in 
astrophysics. Such remnants can be formed during the explosions of type II 
supernovae \cite{BK,Woosly}, during the mergers of close neutron-star 
binaries \cite{Janka1}, and during accretion-driven collapse 
\cite{Spruit1}. About 10\% of the released gravitational energy is known to 
transform into the thermal energy of the remnant and to be emitted in the 
form of neutrinos, irrespective of the origin of the collapse 
\cite{BK,Woosly,Janka1,Spruit1,Raffelt}. A solid rotating core (of a 
protoneutron star or a black hole) and a disk (envelope) rotating with an 
angular-velocity gradient are formed in the remnant of collapse on time 
scales of the order of a second. The compact core of typical sizes  $ R_c 
\sim 10 $ km, subnuclear densities, and high temperatures ($ T \gtrsim 10 
$ Mev) is opaque to neutrinos. Being less dense and hot ($ T \sim 3 - 6 $ 
Mev), the envelope with typical sizes of several tens of km is partially 
transparent to neutrinos \cite{Woosly,Janka1,Spruit1,Raffelt}. Within 
several seconds after its formation, the remnant of collapse effectively 
cools down via neutrino radiation. The neutrino luminosities during this 
period are typically $ L_{\nu}\sim 10^{52} $ erg s$^{-1}$ \cite{Raffelt} .
(the passage stage of the main neutrino flux). This neutrino flux could 
have a strong effect on the dynamics of the envelope, in particular, it 
could trigger its ejection \cite{CW}. Detailed calculations show that for 
spherically symmetric collapse, the neutrino absorption by the envelope is 
not intense enough for its ejection \cite{ImNad}. However, such envelope 
ejection is possible for a millisecond remnant of collapse with a strong 
toroidal magnetic field (the magnetorotational model of a supernova 
explosion \cite{BK}).

Because of its rapid rotation and large viscosity, convection, a turbulent 
dynamo, and large angular velocity gradients arise in the remnant of 
collapse. These processes can lead to the rapid (of the order of several 
seconds) generation of a strong poloidal magnetic field with a strength up 
to $ B \sim 3\cdot 10^{15} $ G on a coherence length $ L \sim 1 $ km in 
the core of the remnant of collapse \cite{DunkThom}. Having arisen in this 
way, the strong poloidal field can persist in a young neutron star for 
about $ \sim 10^{3}- 10^{4} $ years. Duncan and Thompson \cite{DunkThom} 
called such young pulsars magnetars. There are strong grounds for 
suggesting that magnetars are observed in nature as soft gamma-ray 
repeaters (SGRR) \cite{Kouvel} or as anomalous X-ray pulsars (AXP) 
\cite{AXP}. An angular velocity gradient in the envelope leads to the 
generation of a secondary toroidal magnetic field through the winding of 
field lines of the primary poloidal magnetic field frozen in the rotating 
envelope plasma. In this case, strong poloidal magnetic fields of strength 
$ B \sim 10^{15}-10^{17} $ G can emerge in the envelope of a millisecond 
remnant of collapse in typical times of $ \sim 1 $ s \cite{Bisnovat}. Such 
strong fields can significantly affect the envelope dynamics even if they 
persist for several seconds. For example, a magnetic field of strength 
$ B \sim 10^{17} $ G can produce an anisotropic gamma-ray burst 
\cite{KlRud} and, as we already pointed out above, can trigger the 
ejection of the supernova envelope \cite{BK}.

Significantly, such strong magnetic fields can arise in the envelope 
during the passage of the main neutrino flux through it (within several 
seconds after collapse). Therefore, it is of interest to investigate the 
possible dynamical effects that emerge when the neutrino flux passes 
through a strongly magnetized envelope. Effects of this kind are discussed 
in the literature. For example, Bisnovatyi-Kogan \cite{Bisnovat2} 
estimated the magnetar velocity gained through a neutrino dynamical kick 
in the case where the magnetic-field strengths on its hemispheres differed 
significantly. In this case, the kick arises because the energy 
transferred from the neutrinos to the medium in $\beta$-processes depends 
on the magnetic field strength. However, we will discuss the dynamical 
effects of a different nature. Because of the P-parity breakdown in weak 
processes, neutrinos are known to be emitted and absorbed in a magnetic 
field asymmetrically, which can lead to macroscopic momentum transfer from 
the neutrinos to the medium \cite{Chugai,Dorof}. Thus, the envelope region 
filled with a strong toroidal magnetic field can acquire a large 
macroscopic angular momentum when an intense neutron flux passes through 
it \cite{GO}.

Here, we estimate the asymmetry in momentum transfer from the neutrinos to 
the medium during the neutrino interaction with nucleons in the envelope 
of a remnant of collapse at the stage of the main neutrino radiation. This 
paper has the following structure. In Section 2, we stipulate basic 
physical assumptions about the parameters of the medium, magnetic-field 
strengths, and the neutrino distribution function. In Section 3, we derive 
general expressions for the reaction rates and for the components of the 
energy-momentum transferred from the neutrinos to a volume element of the 
medium per unit time in the direct URCA processes for a strong magnetic 
field. In Section 4, the same quantities are calculated for the scattering 
of antineutrinos of all types by nucleons of a magnetized medium. 
Numerical estimates for the parameters of the medium, neutrino mean free 
paths, and the force density along the magnetic field are presented in 
Section 5. In Section 6, we compare our results with the results of 
similar calculations and briefly discuss the possible dynamical effects of 
the P-parity breakdown in the envelope of a collapsing star. The 
calculations of our basic results are detailed in Appendices A, B, and C. 
Throughout the paper, we use a system of units with 
$ c = \hbar = k_B = 1 $.

\section{
Physical assumptions
}

Here, we investigate basic neutrino-nucleon processes in the envelope of a 
remnant of collapse with a strong magnetic field at the stage of the main 
neutrino radiation. We consider the direct URCA processes 
\be
n + \nu_e \Longrightarrow  p + e^- ,
\lab{1} \\
p + e^- \Longrightarrow  n + \nu_e ,
\lab{2} \\
p +  \tilde\nu_e \Longrightarrow  n + e^+,
\lab{3} \\
n + e^+ \Longrightarrow  p + \tilde\nu_e ,
\lab{4}
\ee
and the scattering of neutrinos of all types by nucleons:
\be
&&N + \nu_i \Longrightarrow N + \nu_i  ,
\lab{5} \\
&&N + \tilde\nu_i \Longrightarrow N + \tilde\nu_i ,
\lab{6} \\
&&( \nu_i = \nu_e,  \nu_{\mu},  \nu_{\tau} ),   ( N = n,  p ).
\nonumber
\ee
Note that $\beta$-decay is statistically suppressed under the conditions 
considered. A quantitative estimate of the asymmetry in momentum transfer 
follows from the expression for the energy-momentum transferred in these 
processes from the neutrinos to a unit of volume of the medium per unit 
time: 
\be
\frac{dP_{\alpha}}{dt} 
=
\left(  \frac {dQ}{dt}, {\vec \Im} \right)
 = 
\frac{1}{V} \int \prod\limits_i dn_i f_i
\prod\limits_f dn_f (1-f_f)
\frac{|S_{if}|^2}{{\cal T}} q_{\alpha},
\lab{dpa}
\ee
Here, $ dn_i $ and $ dn_f $ are the numbers of initial and final states in 
the element of phase volume; $ f_i $ and $ f_f $ are the distribution 
functions of the initial and final particles; $ |S_{if}|^2/{\cal T} $ is 
the square of the S-matrix element for the process per unit time; and 
$ q_{\alpha} $ is the 4-momentum transferred to the medium in a single 
reaction. An important quantity that characterizes the process is also its 
rate $ \Gamma $ defined as
\be
\Gamma =  \frac{1}{V} \int \prod\limits_i dn_i f_i
\prod\limits_f dn_f (1-f_f)
\frac{|S_{if}|^2}{{\cal T}}  .
\lab{g}
\ee
In particular, the neutrino mean free paths can be easily determined from 
this quantity: 
\be
\bar l_{\nu} = \frac{N_{\nu}}{\Gamma^{tot}_{\nu}} ,
\lab{lnu}
\ee
where $N_{\nu}$ is the local neutrino number density and 
$\Gamma^{tot}_{\nu}$ is the sum of the absorption and scattering rates for 
the neutrinos of a given type.

In the reactions of neutrino interaction with matter, we separate the 
medium and the neutrino flux passing through it. By the medium, we mean 
free electrons, positrons, and nucleons with an equilibrium Dirac 
distribution function:
\be
f_i
=
\frac {1} { \exp ( E_i / T - \eta_i ) + 1} ,
\nonumber
\ee
where $ \eta_i = \mu_i / T $;  $ E_i $ and $ \mu_i $ are the energy and 
chemical potential for the particles of a given type. We disregard the 
effect of envelope rotation on the distribution functions, because the 
rotation velocity is nonrelativistic even for a millisecond remnant of 
collapse. At typical (for the envelope) densities 
($ \rho \sim 10^{11} - 10^{12} $ g cm$^{-3}$) and temperatures 
($ T \sim $ several MeV), the \ep-plasma is ultrarelativistic and the 
nucleon gas is a Boltzmann nonrelativistic one.

We consider the case of a strong magnetic field, i.e., assume that the 
parameters of the medium and the magnetic field strength are related by 
\be
m_p T \gg 2 eB \gtrsim \mu_e^2, T^2 \gg m_e^2 ,
\lab{eB}
\ee
where $ m_p $ and $ m_e $ are the proton and electron masses and 
$ e > 0 $ is an elementary charge. Condition (\ref{eB}) implies that the 
electrons and positrons occupy mainly the lower Landau level, while the 
protons occupy many levels.

The quantity (\ref{dpa}) is known to be zero \cite{Kusenko} for a thermal 
equilibrium of the neutrinos with the medium. However, we consider the 
envelope region where the neutrino distribution function deviates from the 
equilibrium one. In the model of spherically symmetric collapse, the local 
(anti)neutrino distribution without a magnetic field can be fitted as 
follows \cite{Raffelt,HM}: 
\be
f_{\nu} 
= 
\frac { \Phi_{\nu}(r,\chi) }
{ \exp ( \omega / T_{\nu} - \eta_{\nu} ) + 1 }  .
\lab{fn}
\ee
Here, $ \chi $ is the cosine of the angle between the neutrino
momentum and the radial direction, $ \omega $ is the neutrino energy,  
$ T_{\nu} $ is the neutrino spectral temperature, and $ \eta_{\nu} $ is 
the fitting parameter. In this paper, we ignore the effect of a magnetic 
field on the neutrino distribution function. This approximation is good 
enough when the neutrino mean free path is larger than or of the order the 
region occupied by a strong magnetic field. In the model considered in 
\cite{Bisnovat}, this region does not exceed few kilometers in size, 
whereas the mean free paths for the neutrinos of different types are 
estimated to be 2-5 km. Thus, in our subsequent calculations, we use the 
neutrino distribution function (\ref{fn}). We will return to a discussion 
of this issue in Section 5.

\section{
Direct URCA processes
}

We used the standard low-energy Lagrangian for the weak interaction of 
neutrinos with nucleons to calculate the S-matrix element of the direct 
URCA processes. Our calculations of the square of the S-matrix element are 
detailed in Appendix A. In the limit of a strong magnetic field, where the 
electrons and positrons occupy only the ground Landau level, we derived 
the expression 
\be
&&\frac{|S_{if}|^2}{{\cal T}} = \frac{G_F^2  \cos^2 \theta_c  \pi^3}
{ 2  L_y  L_z  V^2  \omega  \varepsilon}
\exp{(-Q_{\perp}^2 / 2eB )}
\Bigg[
\sum \limits_{n=0}^{\infty}
\frac{|M_+|^2}{n!}
\left( \frac{Q_{\perp}^2}{2eB}
\right)^n
\delta^{(3)} +
\nonumber  \\
&&\sum\limits_{n=1}^{\infty}
\frac{|M_-|^2}{(n-1)!}
\left( \frac{Q_{\perp}^2}{2eB}
\right)^{n-1}
\delta^{(3)} \Bigg] ,
\lab{S} \\
&&|M_+|^2 = 4
\bigg( \varepsilon + p_{\|} \bigg)
\Bigg[
\left( 1 + g_a \right)^2 \bigg( \omega + k_{\|} \bigg)
+ 4 g_a^2 \bigg( \omega - k_{\|} \bigg)
\Bigg]
\lab{M+} , \\
&&|M_-|^2 = 4
\left( 1 - g_a \right)^2
\bigg( \varepsilon + p_{\|} \bigg)
\bigg( \omega + k_{\|} \bigg)
\lab{M-} .
\ee
Here, $\delta^{(3)}$ is the delta function of the energy, the momentum 
along the magnetic field, and one of its transverse components that are 
conserved in the reactions; is the square of the transferred momentum
across the magnetic field in the corresponding reaction; 
$Q_{\perp}^2$ is the index of summation over the proton Landau levels; 
$ \varepsilon $, $ p_{\|} $, $ \omega $, and $ k_{\|} $ are the energy and 
the momentum component along the magnetic field for the electron and 
neutrino, respectively; $ {\cal T} L_x L_y L_z $ is the normalization 
4-volume; $g_a$ is the axial constant of a charged nucleon current 
($ g_a \simeq 1.26 $ in the low-energy limit); $G_F$ is the Fermi 
constant; and $\theta_c$ is the Cabibbo angle.

Note that we derived the above expression for $|S_{if}|^2$ in \cite{GO}. 
It is identical to that from \cite{LA} and \cite{BY} when the erroneous 
sign of the axial constant $g_a$ is corrected (for the formal change 
$ g_a \to - g_a $). However, the paper \cite{LP}, where only the term 
$ n = 0 $ in the square of the $S$-matrix element was calculated, contains 
a difference in expression (\ref{M+}). To obtain the expression for $S^2$ 
from \cite{LP}, the factor $(1+g_a)^2$ must be substituted for $(1+g_a^2)$ 
in the first term of the amplitude (\ref{M+}).

For convenience, we calculate the 4-momentum, $ dP_{\alpha} / dt $, 
transferred to the medium in the direct URCA processes involving neutrinos 
(\ref{1}), (\ref{2}) and antineutrinos (\ref{3}), (\ref{4}) separately.  
Using the $T$ invariance of the square of the $S$-matrix element for these 
processes and explicit particle distribution functions for the medium, we 
can represent the quantity (\ref{dpa}) as 
\be
&&\frac{dP_\alpha^{(\n)}}{dt} 
= 
\int \frac{d^3k}{(2\pi)^3}  k_\alpha 
{\cal K}^{(\n)}
\Bigg[
\Bigg(
1+ \exp{\bigg( -\omega / T \pm \delta \eta \bigg)}
\Bigg)
f_{\n} - \nonumber \\
&&- \exp{\bigg( -\omega / T \pm \delta \eta \bigg)}
\Bigg] , 
\lab{dpurca}
\ee
Here, $ \delta\eta = (\mu_e + \mu_p - \mu_n) / T $; $k_{\alpha}$ is the 
4-momentum of the (anti)neutrino, and ${\cal K}^{(\n)}$ is the absorption 
coefficient in the (anti)neutrino absorption reaction defined as 
\be
{\cal K}^{(\n)} 
=  
\int dn_p  dn_n  dn_e 
\frac{|S_{if}|^2}{{\cal T}} 
\left\{
\begin{array}{cc}
f_n  (1-f_p)  (1-f_{e^-})  \\
f_p  (1-f_n)  (1-f_{e^+})
\end{array}
\right\} .
\lab{K}
\ee
Note that under $\beta$-equilibrium conditions, when 
$ \eta_{\nu} = \delta \eta = (\mu_e + \mu_p - \mu_n) / T $ 
expression (\ref{dpurca}) for the transferred momentum is 
\be
\frac{dP_\alpha^{(\n)}}{dt} = \int \frac{d^3k}{(2\pi)^3}  k_\alpha 
\left(
1 + \exp{
\left(
- \frac{\omega}{T} \pm \eta_{\nu}
\right)}
\right) 
{\cal K}^{(\n)}  \delta f^{\n}  ,
\ee
where $\delta f^{\n}$ is the deviation of the neutrino distribution 
function from the equilibrium one. Thus, the 4-momentum transferred from 
the neutrinos to the medium is nonzero only in a partially transparent 
(to neutrinos) envelope and vanishes in a dense $\beta$-equilibrium 
remnant core.

We calculated the absorption coefficient (\ref{K}) in the strong field 
limit (\ref{eB}) by assuming the nucleon gas to be nondegenerate. In this 
case, the calculation technique is simplified sharply compared to the more 
general case. For this reason, we discuss the details of our calculation 
of the absorption coefficient in Appendix B. Here, we present only the 
final result that was previously published \cite{GO} without explaining 
the details of our calculations: 
\be
{\cal K}^{(\n)} =
\frac{ G_F^2 \cos^2 \theta_c }{ 2\pi }
eB N_{(n,p)}
\Bigg(
\bigg( 1 + 3 g_a^2 \bigg) -
\frac{k_\|}{\omega}  \bigg( g_a^2-1 \bigg)
\Bigg)
\Bigg(
1  +  \exp{ \left( \pm a - \frac{\omega}{T} \right) }
\Bigg)^{-1}
\! ,
\lab{Kn}
\ee
where the terms of order $ \sim  eB / m_p T $ were discarded. 
Here, $N_n$, $N_p$, $m_n$, and $m_p$ are the neutron and proton number 
densities and masses, $ a = (\mu_e - \triangle) / T $, 
$ \triangle = m_n - m_p $. Note that the derived expression is identical 
to that in \cite{LA}.

For completeness, we give the expressions for the energy density 
$ dQ / dt $ and for the rates $\Gamma$ of the direct URCA processes 
(\ref{1}) - (\ref{4}): 
\be
\left\{
\begin{array}{cc}
\Gamma \\
dQ / dt
\end{array}
\right\}_{n +\nu_e \rightarrow p+e^-}  =
{\cal A}  J_\nu  \frac{N_n}{N_B}
\left\{
\begin{array}{cc}
T^3  C_2(a,T_\nu,\eta_\nu) \\
T^4  C_3(a,T_\nu,\eta_\nu)
\end{array}
\right\} ,
\lab{q1}
\ee
\be
\left\{
\begin{array}{cc}
\Gamma \\
dQ / dt
\end{array}
\right\}_{p+e^- \rightarrow n + \nu_e}  =
{\cal A}    \frac{N_p}{N_B}   e^{\triangle / T}
\left\{
\begin{array}{cc}
T^3  B_2(a) - T^3  J_\nu  D_2(a,T_\nu,\eta_\nu)   \\
- T^4  B_3(a) + T^4  J_\nu  D_3(a,T_\nu,\eta_\nu)
\end{array}
\right\}  ,
\lab{q2}
\ee
\be
\left\{
\begin{array}{cc}
\Gamma \\
dQ / dt
\end{array}
\right\}_{p + \tilde \nu_e \rightarrow n + e^+}  =
{\cal A}  J_{\tilde\nu}  \frac{N_p}{N_B}
\left\{
\begin{array}{cc}
T^3  C_2(-a,T_{\tilde\nu},\eta_{\tilde\nu}) \\
T^4  C_3(-a,T_{\tilde\nu},\eta_{\tilde\nu})
\end{array}
\right\}   ,
\lab{q3}
\ee
\be
\left\{
\begin{array}{cc}
\Gamma \\
dQ / dt
\end{array}
\right\}_{n + e^+ \rightarrow p + \tilde \nu_e }  =
{\cal A}    \frac{N_n}{N_B}   e^{-\triangle / T}
\left\{
\begin{array}{cc}
T^3  B_2(-a) - T^3  J_{\tilde\nu} 
D_2(-a,T_{\tilde\nu},\eta_{\tilde\nu})   \\
- T^4  B_3(-a) + T^4  J_{\tilde\nu} 
D_3(-a,T_{\tilde\nu},\eta_{\tilde\nu})
\end{array}
\right\}    ,
\lab{q4}
\ee 
where the dimensional coefficient {\cal A} is defined as 
\be
{\cal A} = \frac{2 G_F^2 \cos^2 \theta_c}{(2\pi)^3}
\bigg( 1 + 3 g_a^2 \bigg)   eB   N_B  ,  
\nonumber 
\\
N_B = N_n + N_p  .  
\nonumber
\ee
Our introduced functions $B_n,  C_n, and D_n$ can be expressed in terms of 
the integrals as follows: 
\be
&&B_n(a) = \int\limits_0^\infty \frac{Z^n dZ}{ e^{Z - a} + 1}  ,
\nonumber  
\\
&&C_n(a,T_\nu,\eta_\nu) =
\int\limits_0^\infty
\frac{Z^n dZ}
{ \bigg( e^{a-Z} + 1 \bigg)
\bigg( e^{ Z  T / T_\nu  - \eta_\nu} + 1 \bigg)}  ,
\nonumber  
\\
&&D_n(a,T_\nu,\eta_\nu) =
\int\limits_0^\infty
\frac{Z^n dZ}
{ \bigg( e^{Z-a} + 1 \bigg)
\bigg( e^{ Z  T / T_\nu  - \eta_\nu} + 1 \bigg)}  .
\nonumber
\ee
The parameter 
\be
J_{\nu} 
= 
\left(
\int f_\nu d^3k
\right)
\left(
\int \Big(
1 + \exp \left( \omega / T -\eta_\nu \right)
\Big)^{-1}
d^3k
\right)^{-1}
\label{J}
\ee 
has the meaning of the ratio of the actual local neutrino number density 
to the equilibrium one with temperature $T_{\nu}$.

Our calculations of the components of momentum (\ref{dpurca}) transferred 
to the medium during neutrino reradiation show that the emerging radial 
force is much weaker than the gravitational force and cannot significantly 
affect the envelope dynamics. Thus, of interest is the force component 
acting along the magnetic field. For a toroidal field, the density of this 
force can be represented as 
\be
&&\Im^{urca}_\| = \frac{1}{6} \frac{g_a^2 - 1}{3g_a^2 + 1} 
{\cal A}    T^4  
\Bigg[
\frac{N_p}{N_B}
\bigg( 3 \langle \chi^2_\nu \rangle - 1 \bigg)
e^{\triangle / T}    B_3(a)  +
\nonumber  
\\
&&\frac{N_n}{N_B}
\bigg( 3 \langle \chi^2_{\tilde\nu} \rangle - 1 \bigg)
e^{-\triangle / T}    B_3(-a)
\Bigg]  -
\nonumber  
\\
&&\frac{1}{2}  \frac{ g_a^2 - 1 }{ 3 g_a^2 + 1 }
\Bigg[
\bigg( 1- \langle \chi^2_\nu \rangle \bigg)
\frac{dQ_\nu}{dt}  +
\bigg( 1- \langle \chi^2_{\tilde\nu} \rangle \bigg)
\frac{dQ_{\tilde\nu}}{dt}
\Bigg]  .
\lab{F}
\ee
Here, 
\be
\langle \chi^2_\nu \rangle 
=
\left(
\int \chi^2 \omega f_\nu d^3k
\right)
\left(
\int \omega f_\nu d^3k
\right)^{-1} ,
\label{chi}
\ee 
where $ \langle \chi^2_\nu \rangle $ is the mean square of the cosine of 
the angle between the neutrino momentum and the radial direction. As 
follows from this expression, the asymmetry in momentum transfer is 
nonzero for two reasons: either the neutrino distribution is anisotropic 
($ \langle \chi^2 \rangle \neq 1 / 3 $) or energy is transferred to the 
medium in the URCA processes (\ref{1}) - (\ref{4}) 
($ dQ_{\n} / dt \neq 0 $). Interestingly, for an isotropic neutrino 
distribution, the force density is directed along the field when the 
medium cools down and opposite to the field when it heats up through the 
URCA processes.

\section{
The neutrino scattering by nucleons
}

We used the nonrelativistic vacuum wave functions of nucleons with a 
certain spin component along the magnetic-field direction to calculate the 
$S$-matrix element in the reactions of (anti)neutrino scattering by 
nucleons (\ref{5}) and (\ref{6}). Our calculations are detailed in 
Appendix C. Below, we give only the final expression for the square of the 
$S$-matrix element for the neutrino scattering by nucleons: 
\be
&& \frac{ |S_{if}|^2_{\nu} }{ {\cal T} } =
\frac{ (2 \pi)^4 G_F^2 }{ 2 V^3 \omega \omega' }  
\delta^{(4)}  
\Bigg[
\left(   c_v^2 + 3 c_a^2    \right) \omega \omega'
+ \left(  c_v^2 - c_a^2  \right) \left( \vec k \vec k' \right) +
\nonumber 
\\
&& + 2 c_v c_a  \left(  \omega k'_\| + \omega'  k_\| \right)
\left(  S + S'  \right)
- 2 c_a^2  \left(   \omega k'_\| - \omega'  k_\|   \right)
\left(  S - S' \right)   +   
\nonumber 
\\
&& +  \left(    c_v^2 - c_a^2    \right)
\left( \omega \omega' + \left( \vec k \vec k'\right) \right) S S'
+ 4 c_a^2 k_\| k'_\|  S S'
\Bigg] .   
\lab{S_sc}
\ee 
The expression for $ |S_{if}|^2 / {\cal T} $ for the antineutrino 
scattering by nucleons can be derived from (\ref{S_sc}) by the change 
$ k \leftrightarrow k' $:
\be
|S_{if}|^2_{\tilde\nu} (k,k') = |S_{if}|^2_{\nu} (k',k) .
\ee
Here, $c_v$ and $c_a$ are the vector and axial constants for a neutral 
nucleon current. In the low-energy limit \cite{RafShekl}, 
\be
&&c_v = - 1 / 2 , \;
c_a \simeq - 0.91 / 2  \;\; {\rm for neutrons},
\nonumber \\
&&c_v  = 0.07 / 2 , \;
c_a \simeq  1.09 / 2   \;\;  {\rm for protons}. 
\lab{cvca}
\ee
The conservation of energy-momentum is defined as 
$\delta^{(4)}\! = \! \delta^{(4)}( {\cal P} + k - {\cal P'} - k' )$, 
where $ k \! = \! (\omega, \vec k) $; $ k' \! = \! (\omega', \vec k') $; 
${\cal P}$ and ${\cal P'}$ are the 4-momenta of the initial and final 
neutrinos and nucleons, respectively;
$k_\|$ and $k'_\|$ are the momentum components of the initial and final 
neutrinos along the magnetic field; and $S$ and $S'$ are the components of 
the polarization vectors for the initial and final nucleons along the 
magnetic field ($ S = \pm 1 $).

Analysis of the kinematics of the neutrino scattering by nucleons shows 
that the energy transferred in these reactions to an element of the medium 
is negligible compared to the energy transferred in the URCA processes. 
This can also be verified by a direct calculation using the technique 
detailed in Appendix C. Therefore, below, we are concerned only with the 
component of the momentum transferred along the magnetic field. This 
component arises from a partial polarization of the nucleon gas in the 
field, because nucleons with different polarizations have different 
energies: 
\be
E_N = m_N + \frac{ \vec {\cal P}^2 }{ 2 m_N } -
g_N  S  \frac{ eB }{ 2 m_N }   , 
\nonumber
\ee
where $g_N$ is the nucleon magnetic factor ($ g_n \simeq  -1.91 $ for the 
neutron and $ g_p \simeq 2.79 $ for the proton). Taking into account the 
energy of interaction between the nucleon magnetic moment and the magnetic 
field, we obtained the following expression for the force density along 
the magnetic field during the scattering of neutrinos of one type by 
neutrons or protons (see Appendix C for the details of our calculations): 
\be
\Im _\| ^{(\nu_i)} = - \frac{ G_F^2 g_N } { 2 \pi }  
\frac{ eB }{ m_N T }   N_N  N_\nu  
\Bigg\{
\bigg( c_v  c_a \langle \omega^3_\nu \rangle +
c_a^2 T \langle \omega^2_\nu \rangle \bigg)
\bigg( \langle \chi^2_\nu \rangle - 1 / 3 \bigg) -
\nonumber \\
- c_a^2 \bigg(
\langle \omega^3_\nu \rangle - 5 T \langle \omega^2_\nu \rangle
\bigg)
\bigg(
5 / 3 -  \langle \chi^2_\nu \rangle
\bigg)
+ 2 c_a^2 J_{\nu}
\bigg(
\langle \omega^3_\nu \rangle -
5T_\nu \langle \omega^2_\nu \rangle
\bigg)
\bigg(
1 - \langle \chi^2_\nu \rangle
\bigg)
\Bigg\} ,
\lab{F3} 
\ee
where 
\be
\langle \omega^n_\nu \rangle 
= 
\left(
\int \omega^n f_\nu d^3k 
\right)
\left(
\int f_\nu d^3k 
\right)^{-1} ,
\label{e}
\ee
$ J_{\nu} $ and $ \langle \chi^2_\nu \rangle $ are the quantities defined 
in the same way as in the URCA processes, and $N_\nu$ is the local neutrino 
number density. During the antineutrino scattering (\ref{6}), the momentum 
transferred to the medium per unit time is given by expression (\ref{F3}) 
with the formal change $ c_a^2 \rightarrow - c_a^2 $:
\be
\Im _\| ^{(\tilde\nu_i)} (c_a^2) = \Im _\| ^{(\nu_i)} ( - c_a^2) .
\lab{F4}
\ee

For the Boltzmann neutrino distribution
\be
f_{\nu} = \Phi_{\nu}(r,\chi) \exp{( - \omega / T_{\nu})} ,
\nonumber
\ee
expression (\ref{F3}) can be simplified:
\be
\Im _\| ^{(\nu_i)} = - \frac{ 6 G_F^2 g_N } { \pi }  
\frac{ eB }{ m_N }   N_N  N_\nu   T_\nu^2
\Bigg\{
4 c_a^2
\bigg(
2 - \langle \chi^2_\nu \rangle
\bigg) + 
\nonumber 
\\
+ 5 T_\nu / T
\bigg[
c_v  c_a
\bigg( \langle \chi^2_\nu \rangle - 1 / 3 \bigg)
- c_a^2 \bigg( 5 / 3 -  \langle \chi^2_\nu \rangle \bigg)
\bigg]
\Bigg\}  .
\lab{FB}
\ee
Note also that in the envelope of the remnant of collapse, the parameters 
of the distribution functions for the neutrinos and antineutrinos of type 
$\mu$ and $\tau$ are virtually identical \cite{Janka2}. This allows a 
simple expression to be written for the total (neutrinos plus 
antineutrinos of a given type) force density along the magnetic field: 
\be
\Im _\| ^{(\nu_i)} = - \frac{ G_F^2 c_v  c_a g_N } { \pi }  
\frac{ eB }{ m_N T }   N_N  N_\nu  
\langle \omega^3_\nu \rangle
\bigg( \langle \chi^2_\nu \rangle - 1 / 3 \bigg) .
\lab{Fsum}
\ee

As can be seen from formulas (\ref{F3}) and (\ref{F4})-(\ref{Fsum}), there 
is an asymmetry in momentum transfer along the magnetic field, as in the 
URCA processes, either for an isotropic neutrino distribution 
($ \langle \chi^2 \rangle \neq 1 / 3 $) or when the neutrino spectral 
temperature differs from the temperature of the medium ($ T_\nu \neq T $). 
Interestingly, the force densities along the field $ \Im _\| ^{(\nu_i)} $ 
during the scattering by neutrons and protons are directed oppositely. 
This follows from the fact that the expression for the force density is 
proportional to the nucleon magnetic factor $g$ (recall that 
$ g_n \simeq  -1.91 $ for the neutron and 
$ g_p \simeq 2.79 $ for the proton). Under actual conditions in the 
envelope of the remnant of collapse, it is commonly assumed that 
$ N_p / N_B~\ll~1 $. We see from expressions (\ref{F3}) and (\ref{F4}) 
that the force density is proportional to the nucleon number density. 
Thus, the scattering by neutrons mainly contribute to the quantity 
$\Im_\|$. For completeness, below, we give expressions for the dominant 
contribution (with the terms 
$ eB / m_N T $ and $ \sqrt{T / m_N } $ disregarded) to the rate 
(\ref{g}) of scattering of (anti)neutrinos of any type by nucleons: 
\be
\Gamma^{sc} = \frac{G_F^2}{\pi}  \bigg( c_v^2 + 3 c_a^2 \bigg)  
N_N  N_{\nu}    \langle \omega^2_\nu \rangle 
\Bigg[
1 - J_\nu
\left(
1 - \frac{ 4 T_\nu  \langle \omega_\nu \rangle }
{  \langle \omega^2_\nu \rangle }
\right)
\Bigg]  . 
\lab{G_sc}
\ee
Recall that $ c_v $ and $ c_a $ are the vector and axial constants of a 
neutral nucleon current, which change only when the type of nucleon 
changes [see (\ref{cvca})] in the scattering reactions. Note that the 
scattering reaction rates and, hence, the neutrino mean free paths in 
these processes are virtually independent of the magnetic field.

\section{
Numerical estimates
}

For the force density along the magnetic field generated in the processes 
of neutrino interaction with nucleons (\ref{1})-(\ref{6}) to be 
numerically estimated, the parameters of the medium and the neutrino 
radiation must be specified. Recall that we consider a sufficiently dense 
($ \rho \sim 10^{11} - 10^{12} $ g cm$^{-3}$) and hot 
($ T \sim $ several MeV) envelope region that is partially transparent to 
neutrinos. Part of this envelope, about 1-3 km in size, is assumed to be 
filled with a strong toroidal magnetic field of typical strength 
$ B \gtrsim 10^{16} $ G. For such envelope parameters and magnetic-field 
strengths, condition (\ref{eB}) is satisfied and, hence, all our 
expressions are valid. Recall that we used the local nonequilibrium 
neutrino distribution function (\ref{fn}) while disregarding the effect of 
magnetic field on it. Below, we verify the validity of this approximation 
by estimating the mean free paths for the neutrinos of various types and 
by comparing them with the size of the region filled with a strong 
magnetic field. The results from \cite{Janka2} were used to fit the 
neutrino-radiation parameters. In the above paper, a numerical solution 
was obtained for the neutrino distribution function at the stage of the 
main neutrino radiation after spherically symmetric collapse.

Analysis of the numerical values for $ \langle \omega_\nu \rangle $ and 
$ \langle \omega^2_\nu \rangle $ from \cite{Janka2} shows that the 
distribution functions for the neutrinos of various types are well fitted 
by the parameters: 
\be
&&T_{\nu_e} \simeq 3.3 \; {\rm MeV}, \;\;
T_{\tilde\nu_e} \simeq 4.5 \; {\rm MeV}, \;\;
T_{\nu_{\mu,\tau}}  \simeq  T_{\tilde\nu_{\mu,\tau}} 
\simeq  6.6 \; {\rm MeV},
\nonumber 
\\
&&\eta_{\nu_e} \simeq 2.8 ,  \;\;
\eta_{\tilde\nu_e} \simeq 2.0 , \;\;
\eta_{\nu_\mu,\nu_\tau} \simeq
\eta_{\tilde\nu_\mu,\tilde\nu_\tau} \simeq  2.2   .
\lab{Tn}
\ee
For numerical estimates, we chose an envelope region with a typical 
density of the medium $ \rho \simeq 5 \times 10^{11} $ g cm$^{-3}$ and 
assumed the magnetic-field strength in it to be 
$B \simeq 4 \cdot 10^{16} $ G. The local neutrino number densities in this 
region are 
\be
N_{\nu_e} \simeq 9 \cdot 10^{32} \; {\rm cm}^{-3} , \;\;
N_{\tilde\nu_e} \simeq  3 \cdot 10^{32} \; {\rm cm}^{-3} ,  \;\;
N_{\nu_{\mu,\tau}} \simeq  N_{\tilde\nu_{\mu,\tau}}  \simeq 
2.7 \cdot 10^{32}  \; {\rm cm}^{-3} .
\lab{Nn}
\ee
The following values of our introduced parameters $J_\nu$ correspond to 
these number densities: 
\be
J_{\nu_e} \simeq 0.23 , \;\;
J_{\tilde\nu_e} \simeq 0.05 , \;\;
J_{\nu_{\mu,\tau}} \simeq J_{\tilde\nu_{\mu,\tau}} \simeq 0.01  .
\lab{Jn}
\ee
Accordingly, the mean square of the cosine of the angle between the 
neutrino momentum and the radial direction is 
\be
\langle \chi^2_{\nu_i} \rangle  \simeq 
\langle \chi^2_{\tilde \nu_i} \rangle   \simeq  0.385  .
\lab{chin}
\ee
At the stage of the main neutrino radiation (within about 1-3 s after 
collapse), the parameters of the medium vary slowly through hydrodynamic 
processes compared to the time scales of $ \sim 10^{-2} - 10^{-3} $ s on 
which a quasi-equilibrium is established through the dominant URCA 
processes. For this reason, the medium is assumed to be in 
quasi-equilibrium: 
\be
&&\Gamma_{n \to p} = \Gamma_{p \to n} ,
\lab{eqG}  \\
&&\frac{dQ}{dt} = 0 ,
\lab{eqQ}
\ee 
where $ \Gamma_{n \to p} $ and $ \Gamma_{p \to n} $ are the sums of the 
rates of the processes with the conversion of a neutron into a proton and 
a proton into a neutron, respectively; and $ dQ / dt $ is the total amount 
of energy transferred in the neutrino processes to a unit volume of the 
medium per unit time. Recall that under the envelope conditions 
considered, the URCA processes give a dominant contribution to the 
establishment of equilibrium. Together with the electroneutrality 
condition, 
\be
\frac{N_p}{N_B} = \frac{ eB \mu_e}{ 2 \pi^2 N_B} ,
\lab{elec}
\ee
the quasi-equilibrium equations for the medium (\ref{eqG}) and (\ref{eqQ}) 
allow only the density of the medium and the magnetic-field strength to be 
considered as free parameters. Numerically solving this system of 
equations yields 
\be
T \simeq 3.8 \; {\rm MeV},  \;\;
a \simeq 2.8 , \;\;
\frac{N_p}{N_B} \simeq 0.07 .
\lab{param}
\ee
For these parameters, the neutrino mean free paths (\ref{lnu}) are 
estimated to be 
\be
\bar l_{\nu_e} \simeq 3 \; {\rm km}  ,  \;\;
\bar l_{\tilde\nu_e} \simeq 5 \; {\rm km}  ,  \;\;
\bar l_{\nu_\mu,\nu_\tau} \simeq
\bar l_{\tilde\nu_\mu,\tilde\nu_\tau} \simeq 2.5 \; {\rm km} .
\ee
Comparison of the neutrino mean free paths with the size of a region, 
$\sim $1-3 km, filled with a strong magnetic field shows that the magnetic 
field cannot significantly affect the neutrino distribution functions.

Note that the neutrino scattering by nucleons gives a contribution to the 
$\nu_e$ mean free path comparable to the URCA processes and a dominant 
contribution to the $\tilde\nu_e$ mean free path. Thus, the 
$\tilde\nu_e$, $\nu_\mu$, and $\nu_\tau$ mean free paths are virtually 
independent of the magnetic-field strength.

It is of interest to compare the contributions to the momentum transfer 
along the magnetic field from the URCA processes and from the neutrino 
scattering by nucleons. The scattering of $\mu$ and $\tau$ (anti)neutrinos 
by  neutrons [see (33)] gives a dominant contribution to the 
force density in reactions (5) and (6), whereas the reactions
with electron neutrinos [see formula (24) under
the quasi-equilibrium conditions) give a dominant contribution
in processes (1)-(4). Therefore, the ratio of the
force densities can be represented as
\be
\frac { \Im_\|^{(scat)} } { \Im^{(urca)}_{\|} } 
\simeq 
\frac { 4 c_v c_a g_n } { g_a^2 - 1} 
\frac { B_5 (\eta_{\nu_x}) } { B_3 (a) }
\frac { J_{\nu_x} } {Y} 
\frac { T_{\nu_x} } {m_n} 
\left( 
\frac { T_{\nu_x} } {T} 
\right)^5 ,
\ee 
where $ \nu_x = \nu_\mu $, $ \nu_\tau $, and 
$ Y = N_p / ( N_p + N_n ) $ is the chemical composition parameter of the 
medium. By substituting numerical parameters of the medium and the 
neutrinos in this expression, we can easily verify that it is of the order 
of unity, although the $ T_{\nu_x} / m_n $ ratio is small.

Under the quasi-equilibrium conditions for the medium 
(\ref{eqG}) and (\ref{eqQ}), the expression for the force density along 
the magnetic field in the URCA processes (\ref{F}) is greatly simplified. 
Its numerical estimate for the above parameters is 
\be
\Im^{(urca)}_{\|} 
\simeq 
1.4 
\times 10^{20}  
\; {\rm \frac{dyne}{cm^3}}
\left( \frac{B}{ 4.4 \times 10^{16}  \; {\rm G} } \right)  .
\lab{FF1}
\ee 
As expected, the scattering of neutrinos of all types by neutrons gives a 
numerically larger estimate for the total [over all types of 
(anti)neutrinos] force density than do the URCA processes:
\begin{eqnarray}
\Im_\|^{(scat)} 
\simeq  
3.0 
\times 10^{20} 
\; {\rm \frac{dyne}{cm^3}}
\left( \frac{B}{ 4.4 \times 10^{16} \; {\rm G} } \right)
\left( \frac{ \rho }{ 5 \times 10^{11}  \;  {\rm g \; cm^{-3}} }
\right) .
\lab{FF2}
\end{eqnarray}
We numerically analyzed the quasi-equilibrium conditions in the envelope 
region specified by the range of densities 
$ 2 \times 10^{11} \le \rho \le 10^{12} $ g cm$^{-3}$ with the neutrino 
parameters at a given density from \cite{Janka2}. Our analysis shows that 
the force density changes smoothly with increasing density of the medium. 
In the scattering processes, the force density increases almost linearly; 
in the URCA processes, the monotonic rise gives way to a decrease at 
$ \rho \simeq 8 \cdot 10^{11} $ g cm$^{-3}$. The value of 
$ \rho = 5 \times 10^{11} $ g cm$^{-3}$ that we chose for our estimates is 
actually a point with a mean force density in the above range of 
densities. Note that the force generated in the two processes is directed 
along the field (i.e., the effect in all processes of neutrino interaction 
with nucleons is accumulated) and is large quantitatively. Interestingly, 
under the quasi-equilibrium conditions, the force density is completely 
and mainly determined by anisotropy of the angular neutrino distribution 
function in the URCA processes and in the scattering process, 
respectively.

To discuss the possible macroscopic effects of neutrino radiation on the 
magnetized medium of the envelope, below, we give an estimate of the 
angular acceleration that arises in an envelope element under the total 
force density of neutrino spinup: 
\begin{eqnarray}
\dot \Omega \sim 10^{3} \; {\rm s^{-2}}
\left( \frac{B}{ 4.4 \times 10^{16} \; {\rm G}} \right)
\left( \frac{ R }{ 10 \; {\rm km} } \right) ,
\lab{omega}
\end{eqnarray}
where $R$ is the distance from the envelope volume to the center of the 
remnant of collapse. To obtain this estimate, we assumed that the 
macroscopic momentum is transferred to the entire envelope element. Note 
that this angular acceleration is large enough to spin up the envelope 
region filled with a strong magnetic field to millisecond rotation periods 
in a time of $\sim$1 s.

\section{
Discussion
}

Here, we investigate the possible dynamical effects of P-parity breakdown 
during the neutrino interaction with nucleons in the envelope of a 
collapsing star with a strong magnetic field. These effects are known to 
disappear in an optically thick (for neutrinos) medium \cite{LA}, where 
the neutrino mean free path is much smaller than the characteristic size 
of the envelope of the remnant of collapse (see Section 5). This 
determines the range of envelope densities and temperatures under 
consideration.

Another important question is: At what strengths of the magnetic fields 
might their significant effect on the processes under consideration be 
expected? Currently, the field effect on individual neutrino-nucleon 
processes are being intensively studied. In particular, Leinson and Perez 
\cite{LP} calculated the neutrino luminosity in the direct URCA processes 
in the core of a neutron star (a strongly degenerate nucleon medium with a 
typical density $ \rho > \rho_{nucl} = 2.8 \times 10^{14} $ g cm$^{-3}$ 
and temperature $ T \sim 0.1 $ MeV). It is argued that the magnetic field 
significantly affects the luminosity only upon reaching an enormous 
strength $ B \gtrsim 7 \cdot 10^{17} $ G. Baiko and Yakovlev \cite{BY} 
showed that the direct URCA processes could have neutrino luminosities 
higher than the modified URCA processes at field strength 
$ B \gtrsim 10^{16} $ G. Calculations of the Rosseland mean absorption 
coefficients for neutrinos in the envelope of a collapsing star indicate 
that fields $ B \gtrsim 4 \cdot 10^{15} $ G are required to change these 
coefficients at least by 5\% \cite{LQ}. As a continuation of these 
studies, Lai and Arras \cite{LA} calculated the collision integral for 
basic neutrino-nucleon processes in the Boltzmann equation for the 
neutrino distribution function and its first moments. Estimates of the 
P-odd terms in the moments of the collision integral show that the effects 
of asymmetry in momentum transfer along the magnetic field in the envelope 
of a collapsing star can be significant only when strengths 
$ B \gtrsim 10^{16} $ G are reached. The emergence of such (in order of 
magnitude) magnetic fields when estimating the P-odd effects is natural. 
Indeed, only the electrons and positrons at the ground Landau level 
contribute to the P-odd part of the absorption coefficients for the direct 
URCA processes. The significant dynamical effects related to the P-parity 
breakdown take place at a high \ep-plasma density at the ground Landau 
level, which is ensured by conditions (\ref{eB}).

Assuming that the plasma electrons and positrons are only at the ground 
Landau level, we derived simple analytic expressions for the force density 
along the magnetic field (\ref{F}), rate, and energy transferred to a unit 
volume of the medium per unit time (\ref{q1}) - (\ref{q4}) in the URCA 
processes, as well as general expressions for the force density (\ref{F3}) 
and the rate (\ref{G_sc}) of scattering of neutrinos of any type by 
nucleons. The asymmetry in momentum transfer along the field in these 
processes is nonzero only in the envelope region partially transparent to 
neutrinos ($ \langle \chi^2 \rangle \neq 1 / 3 $, $ T_\nu \neq T $ ).. 
Assuming that the medium is in quasi-equilibrium via the dominant URCA 
processes specified by Eqs. (\ref{eqG}) and (\ref{eqQ}), we find the 
equilibrium parameters of the medium. Numerical estimates of the force 
density along the field (\ref{FF1}) and (\ref{FF2}) for the equilibrium 
parameters of the medium show that in the sum of the processes of neutrino 
interaction with nucleons, the asymmetry in momentum transfer is 
accumulated and this asymmetry is quantitatively large (\ref{omega}).

It makes sense to compare our estimates with calculations of the same 
quantities in the processes of neutrino interaction with a strongly 
magnetized \ep-plasma \cite{KM}. It is important to note that the force 
density along the field in these process is directed along the field and 
may be of the same order of magnitude as that in the neutrino-nucleon 
processes. Thus, the asymmetry in momentum transfer along the field is 
accumulated through the P-parity breakdown in all significant processes of 
neutrino interaction with the medium. Note, however, that Kuznetsov and 
Mikheev \cite{KM} worked with an equilibrium neutrino distribution 
function (a Fermi-Dirac distribution with a spectral temperature 
$T_{\nu}$). This causes the number of neutrino states to be significantly 
overestimated under actual conditions in an envelope partially transparent 
to neutrinos. Indeed, our introduced parameter $J_\nu$, which has the 
meaning of the ratio of the actual neutrino number density to the 
equilibrium one, is much less than unity in the envelope region with the 
densities and temperatures under consideration [see estimate (\ref{Jn})]. 
Thus, in this envelope region, the force density in the processes of 
neutrino interaction with a magnetized \ep-plasma is low compared to its 
value in the processes of neutrino interaction with nucleons.

An asymmetry in momentum transfer along a toroidal magnetic field gives 
rise to the angular acceleration (\ref{omega}) of the envelope region 
filled with such a strong field. This acceleration is large enough to spin 
up this envelope region to millisecond rotation periods in a time of 
$\sim 1 $ s. This large (in magnitude) and local (coordinate-dependent) 
angular acceleration can cause a rapid change in the gradient of angular 
velocities in the envelope, which, in turn, can change the mechanism of 
subsequent generation of a toroidal magnetic field. Indeed, in the 
presence of an additional angular acceleration in the envelope, the 
toroidal magnetic field can grow with time much faster than a linear law 
\cite{Bisnovat}. This toroidal-field rearrangement can affect the 
mechanisms of supernova envelope ejection \cite{BK} and the formation of 
an anisotropic gamma-ray burst in a "frustrated" supernova \cite{PWF}; it 
can also trigger the growth of MHD instabilities \cite{Spruit2}. However, 
to investigate the effect of neutrino spinup on the envelope dynamics and 
the generation mechanism of a toroidal field requires analyzing a complete 
system of MHD equations. This is a complex problem far outside the scope 
of this paper. We hope that it will find its researches.

\section{
Acknowledgments 
}

We wish to thank N.V.~Mikheev for a discussion of virtually all 
fundamental questions of this paper, S.I.~Blinnikov for interest and 
advice on important points of our study. We are also grateful to 
G.S.~Bisnovatyi-Kogan, V.M.~Lipunov, V.B.~Semikoz, M.E.~Prokhorov, 
M.V.~Chistyakov, and participants of the Moscow Workshop of astrophysicists 
(Sternberg Astronomical Institute, Moscow State University) for valuable 
remarks. This work was supported by the Russian Foundation for Basic 
Research (project no. 01-02- 17334) and the Education Ministry of Russia 
(project no. E0011-11.0-5).

\appendixes

\section{
Calculating $|S|^2$ for the Direct URCA Processes
}

In the low-energy limit, the effective $S$-matrix element of the URCA 
processes (\ref{1})-(\ref{4}) can be represented as \cite{Raffelt} 
\be
S_{if} = \frac{G_F \cos \theta_c }{\sqrt{2}}  \int  d^4x \,
\left
[\bar\Psi_{p \sigma}^{(n)}
\gamma_{\alpha} (1 + g_a\gamma_5 ) \Psi_{n \sigma'}
\right] 
\left[
\bar\Psi_e  \gamma_{\alpha} (1 + \gamma_5 ) \Psi_{\nu}
\right]  
\ee
(we use the $\gamma$-matrices in spinor representation with a different 
sign of $\gamma_5$ \cite{PB}). Here, 
$\Psi_e$, $\Psi_{\nu}$, $\Psi_{n \sigma'}$, and $\Psi_{p \sigma}^{(n)}$ 
are the wave functions of the electron, neutrino, neutron, and proton (at 
the nth Landau level), respectively; $ \theta_c $ is the Cabibbo angle; 
$ g_a $ is the axial constant of a charged nucleon current 
($ g_a \simeq 1.26 $ in the low-energy limit); and 
$\sigma$ and $\sigma $ are the components of the double proton and neutron 
spin along the magnetic field, respectively. We performed our calculations 
in a coordinate system where $ {\vec B}=(0,0,B) $ in the gauge 
$ A_{\mu}=(0,0,Bx,0) $. When the \ep-plasma occupies only the ground 
Landau level, the wave functions of relativistic electrons can be chosen 
in the form 
\be
&&\Psi_e =
\frac{ \exp{ [-i (\varepsilon t - p_2 y - p_3 z)] }  
\chi_0 ( \eta^{'} ) }
{ \sqrt{ 2 \varepsilon L_y L_z } }  U_e,  
\eta^{'} = \sqrt{eB}
\left( x + \frac{p_2}{eB} \right) ,
\nonumber 
\\
&&U_e = \sqrt{\varepsilon + m_e}
\left (
\begin{array}{rr}
U 
\\
- p_3 / (\varepsilon + m_e)  U
\end{array}
\right ) , 
U = {0 \choose 1} , 
\ee 
where $ p = (\varepsilon, p_1, p_2, p_3) $ is the electron 4-momentum and 
$ \varepsilon = \sqrt{p_3^2 + m_e^2} $ is its energy. In what follows, 
\be
\chi_n (x) =  \frac{ (eB)^{1/4} e^{ - \eta^2 / 2}}
{ \sqrt{ 2^n n! \sqrt{\pi} }  } 
H_n (x) , 
\ee
$ H_n(x) $ are the Hermitean polynomials, 
$ V = L_x L_y L_z $ is the normalization volume.

The wave function of the nonrelativistic protons at the $n$th Landau level 
was chosen in the form  
\be
&&\Psi_{p \sigma}^{(n)} =
\frac{\exp{ [ - i (E_p t - P_2 y - P_3 z) ] } }
{\sqrt{2 E_p L_y L_z}} 
\Phi_{\sigma}^{(n)} , 
\eta = \sqrt{eB} \left( x - \frac{P_2}{eB}\right) ,
\nonumber 
\\
&&\Phi_{\sigma}^{(n)} = \sqrt{E_p + m_p} {V_\sigma \choose 0}, 
V_{\sigma=+1} = {\chi_n(\eta) \choose 0}, 
V_{\sigma=-1} = {0 \choose \chi_{n-1}(\eta)} ,
\ee
where $ P = (E_p, P_1, P_2, P_3) $ is the proton 4-momentum, 
$ E_p = m_p + P_3^2 / 2 m_p + e B n / m_p $ is the nonrelativistic energy 
of a charge particle at the $n$th Landau level. The wave function for 
neutrons is 
\be
&&\Psi_{n \sigma'}
= \frac{ \exp{ [ -i (E_n t - q_1 x - q_2 y - q_3 z) ] }  }
{ \sqrt{2 E_n V} } 
\sqrt{E_n + m_n} 
{W_{\sigma'} \choose 0} ,
\nonumber 
\\
&&W_{\sigma'}^{+}  W_{\sigma'} = 1 ,
\ee
where $ q = (E_n, q_1, q_2, q_3) $ is the neutron 4-momentum, 
$ E_n = m_n + (\vec q)^2 / 2 m_n $. 
We assume the neutrino to be a standard massless Dirac particle of 
left-handed helicity with the wave function \cite{PB} 
\be
&&\Psi_{\nu} =
\frac{ \exp { [ -i (\omega t - k_1 x - k_2 y - k_3 z) ] } }
{ \sqrt{ 2 \omega V} }
\sqrt{\omega}
{W \choose - W },  
\\
&&W^+ W = 1 ,
\nonumber
\ee
where $ k = ( \omega, k_1, k_2, k_3 ) $ is the neutrino 4-momentum. 
Squaring $ S_{if} $, integrating over $d^4x$, and summing over the 
particle Landau levels and polarizations yields 
\be
\frac{|S_{if}|^2}{{\cal T}} &=& \frac{G_F^2  \cos^2 \theta_c  \pi^3}
{ 2  L_y  L_z  V^2  \omega  \varepsilon}
\exp{(-Q_{\perp}^2 / 2eB )}  \times
\nonumber  
\\
&\times&
\Bigg[
\sum \limits_{n=0}^{\infty}
\frac{|M_+|^2}{n!}
\left( \frac{Q_{\perp}^2}{2eB}
\right)^n
\delta^{(3)} +
\sum\limits_{n=1}^{\infty}
\frac{|M_-|^2}{(n-1)!}
\left( \frac{Q_{\perp}^2}{2eB}
\right)^{n-1}
\delta^{(3)} \Bigg] ,
\ee
Here, $\delta^{(3)}$ is the delta function of the energy, momentum, along 
the magnetic field, and one of its transverse components conserved in the 
reactions; $Q_{\perp}^2$ is the square of the transferred momentum across 
the magnetic field in the corresponding reaction; and $n$ is the index of 
summation over the proton Landau levels. The quantities 
$ |M_+|^2 $ and $ |M_-|^2 $ are defined as 
\be
&&|M_\sigma|^2 = \frac{1}{4 m_p m_n} \sum\limits_{\sigma' = - 1}^{1}
L_{\alpha\beta} N^{\sigma \sigma'}_{\alpha \beta} ,
\label{M}
\\
&&L_{\alpha\beta} = Sp
\Big[
\rho_\nu \gamma_\alpha ( 1 + \gamma_5 ) \rho_e \gamma_\beta
( 1 + \gamma_5 )
\Big] ,
\label{L}
\\
&&N^{\sigma \sigma'}_{\alpha \beta} =
Sp\Big[
\rho_{n \sigma'} \gamma_\alpha ( 1 + g_a \gamma_5 )
\rho_{p \sigma} \gamma_\beta ( 1 + g_a \gamma_5 )
\Big] .
\label{N}
\ee 
Below, we give the neutrino, electron, neutron, and proton density 
matrices that correspond to the wave functions: 
\be
&&\rho_{\nu} = \hat k ( 1 - \gamma_5 ) / 2 , 
\rho_e = ( \hat p_\| + m_e ) \Pi_-  ,
\nonumber \\
&&\rho_{n \sigma'} = m_n ( 1 + \hat u ) \Pi_{\sigma'} , 
\rho_{p \sigma} = m_p ( 1 + \hat u ) \Pi_\sigma ,
\\
&&\hat k = k_\alpha \gamma_\alpha , 
\hat u = u_\alpha \gamma_\alpha , 
\hat p_\| = p_0 \gamma_0 - p_3 \gamma_3 .
\nonumber \\
&&\Pi_\sigma = ( 1 + \sigma i \gamma_1 \gamma_2 ) / 2
\nonumber
\ee
Here, $ \Pi_\sigma $ is the projection operator of the density matrix for 
a charged fermion in a magnetic field (recall that 
$~\sigma,~\sigma'~=~\pm~1~$ are the components of the double proton and 
neutron spin along the magnetic field). To ensure the covariance of our 
calculations, we inserted $ u_\mu = (1, 0, 0, \upsilon) $, the 4-velocity 
of the medium along the field, in the nucleon density matrix (at the end 
of our calculations, we assume that $ \upsilon = 0 $).

In the subsequent calculations, we used the following properties of the 
projection operator: 
\be
\Pi^2_{\sigma} = \Pi_{\sigma} , 
\Pi_\sigma \Pi_{-\sigma} = 0 , 
\Pi_{\sigma} \gamma_{\alpha\|} = \gamma_{\alpha\|} \Pi_{\sigma} , 
\Pi_{\sigma} \gamma_{\alpha\perp} = \gamma_{\alpha\perp} \Pi_{-\sigma} , 
\Pi_{\sigma} \gamma_5 = \gamma_5 \Pi_{\sigma}, 
\label{Ppr}
\ee
where $\gamma_{\alpha\|} = \gamma_{1,2}$ and 
$\gamma_{\alpha\perp} = \gamma_{0,3}$. Since the velocity vector v of the 
medium has only longitudinal components, the nucleon spur (\ref{N}) can be 
reduced to 
\be
N^{\sigma \sigma'}_{\alpha \beta} &=&
m_p m_n  
Sp \left[ 
\Pi_{\sigma'} \gamma_\alpha \Pi_\sigma \gamma_\beta ( 1 - g_a^2 )
\right] 
\nonumber 
\\
&+& m_p m_n 
Sp \left[ 
\Pi_{\sigma'} \gamma_\alpha \Pi_\sigma \hat u_\| \gamma_\beta \hat u_\| 
( 1 + g_a^2 - 2 g_a \gamma_5)
\right]  . 
\ee
In the subsequent calculation of this expression, it is convenient to 
separate out the contributions from identical and different nucleon 
polarizations. The matrices $ \gamma_\alpha $ and $ \gamma_\beta $ have 
only longitudinal components ($ \gamma_\alpha = \gamma_{\alpha \perp} $, 
$ \gamma_\beta = \gamma _{\beta \perp} $) when $ \sigma = \sigma' $ 
and only transverse components ($ \gamma_\alpha = \gamma_{\alpha \perp} $, 
$ \gamma_\beta = \gamma _{\beta \perp} $) when $ \sigma = - \sigma' $. 
This is easy to obtain from the properties of the polarization operator 
(\ref{Ppr}). To calculate the nucleon spur requires the expressions 
\be
&&Sp \Big[ 
\gamma_{\alpha \|} \gamma_{\beta \|} \Pi_{\sigma}
\Big] = 
2 \tilde\Lambda_{\alpha \beta}  , 
\label{Sp1}
\\
&&Sp \Big[ 
\gamma_{\alpha \|} \gamma_{\beta \|} \gamma_5 \Pi_{\sigma}
\Big] = 
2 \sigma \tilde\varphi_{\alpha \beta}  , 
\label{Sp2}
\\ 
&&Sp \Big[ 
\gamma_{\mu \|} \gamma_{\nu \|} \gamma_{\rho \|} \gamma_{\delta \|} \Pi_{\sigma}
\Big] = 
2 \big( 
\tilde\Lambda_{\mu \nu} \tilde\Lambda_{\rho \delta} 
+ \tilde\Lambda_{\mu \delta} \tilde\Lambda_{\nu \rho} 
- \tilde\Lambda_{\mu \rho} \tilde\Lambda_{\nu \delta}
\big) , 
\label{Sp3}
\\ 
&&Sp \Big[ 
\gamma_{\mu \|} \gamma_{\nu \|} \gamma_{\rho \|} \gamma_{\delta \|} 
\gamma_5 \Pi_{\sigma}
\Big] = 
2 \sigma \big(
\tilde\Lambda_{\mu \nu} \tilde\varphi_{\rho \delta} 
+ \tilde\Lambda_{\rho \delta} \tilde\varphi_{\mu \nu}
\big) , 
\label{Sp4}
\\ 
&&\Pi_{\sigma} \gamma_{\alpha \perp} \gamma_{\beta \perp} \Pi_{\sigma} = 
- ( \Lambda_{\alpha \beta} - i \sigma \varphi_{\alpha \beta} ) \Pi_{\sigma} . 
\label{Sp5}
\ee
In what follows, $ \Lambda_{\mu \nu} = ( \varphi \varphi )_{\mu \nu} $ and 
$ \tilde\Lambda_{\mu \nu} = ( \tilde\varphi \tilde\varphi )_{\mu\nu} $,  
where $ \varphi_{\mu \nu} = F_{\mu \nu} / B $ and 
$ \tilde\varphi_{\mu \nu} = \tilde F_{\mu\nu} / B $ 
are the tensor and dual tensor of the external electromagnetic field 
reduced to dimensionless form. The calculated nucleon spur can be reduced 
to 
\be
\left.
N^{\sigma \sigma'}_{\alpha \beta} 
\right|_{\sigma=-\sigma'} 
\!\!\!\!\!\!\!\!\!
= 2 m_p m_n 
\left(
\Lambda_{\alpha \beta} + i \sigma \varphi_{\alpha \beta} 
\right)  
\Bigg[ 
( g_a^2 - 1 ) + ( 1 + g_a^2 ) (u \tilde\Lambda u) 
\Bigg]  ,  
\label{N-}
\\
\left.
N^{\sigma \sigma'}_{\alpha \beta} 
\right|_{\sigma=\sigma'} 
\!\!\!\!\!\!\!\!\!
= 2 m_p m_n 
\Bigg[
( 1 - g_a^2 ) \tilde\Lambda_{\alpha \beta} 
+ ( 1 + g_a^2 ) 
\left(
  2 (u \tilde\Lambda)_\alpha (u \tilde\Lambda)_\beta 
  - u^2 \tilde\Lambda_{\alpha \beta}
\right)  
\nonumber 
\\
- 2 g_a \sigma 
\left(
  (u \tilde\Lambda)_\alpha (\tilde\varphi u)_\beta 
  + (\tilde\varphi u)_\alpha (u \tilde\Lambda)_\beta
\right)
\Bigg]  . 
\label{N+}
\ee
Since tensors (\ref{N+}) and (\ref{N-}) have only longitudinal and 
transverse components, respectively, it is convenient to break up the 
lepton spur into the same structures. When it is convolved with the 
nucleon spur, only the completely longitudinal and completely transverse 
parts of the lepton spur will give a nonzero answer: 
\be
L_{\alpha\beta} N^{\sigma \sigma'}_{\alpha \beta} 
= 
\left. N^{\sigma \sigma'}_{\alpha \beta} \right|_{\sigma=\sigma'}  
L_{\alpha \beta}^{\|}
+
\left. N^{\sigma \sigma'}_{\alpha \beta} \right|_{\sigma=-\sigma'}  
L_{\alpha \beta}^{\perp}
\ee
After simple transformations using the properties of the polarization 
operator (\ref{Ppr}), the corresponding constructions can be reduced to 
\be
L_{\alpha \beta}^{\|} 
=
2 Sp \Big[ 
\gamma_{\beta \|} \hat k_\| \gamma_{\alpha \|} \hat p_\| (1-\gamma_5) \Pi_-
\Big]  ,
\\
L_{\alpha \beta}^{\perp} 
=
2 Sp \Big[ 
\gamma_{\beta \perp} \hat k_\| \gamma_{\alpha \perp} \hat p_\| (1-\gamma_5) \Pi_-
\Big]   .
\ee
Using the above properties (\ref{Sp1})-(\ref{Sp5}), these expressions can 
be easily calculated and represented as 
\be
L_{\alpha \beta}^{\|} 
=
4 \Big( 
(p \tilde\Lambda)_\alpha (k \tilde\Lambda)_\beta 
  + (k \tilde\Lambda)_\alpha (p \tilde\Lambda)_\beta 
  - \tilde\Lambda_{\alpha \beta} (p \tilde\Lambda k)
\Big) 
\nonumber
\\
+ 4 \Big( 
(\tilde\varphi p)_\alpha (k \tilde\Lambda)_\beta 
+ (p \tilde\Lambda)_\alpha (\tilde\varphi k)_\beta
\Big) , 
\\
L_{\alpha \beta}^{\perp} 
=
4 \Big( 
\Lambda_{\alpha \beta} - i \varphi_{\alpha \beta}
\Big) 
\Big( 
(k \tilde\Lambda p) + (k \tilde\varphi p)
\Big) .
\ee
For completeness, we present the convolution of the nucleon and lepton 
spurs with the origin in covariant form: 
\be
\left. N^{\sigma \sigma'}_{\alpha \beta} \right|_{\sigma=\sigma'}  
L_{\alpha \beta}^{\|}
=
8 m_p m_n \Big[ 
2 (1+g_a^2) \left( (u p)_\| (u k)_\| \right) 
+ 4 g_a \sigma (u \tilde\varphi p) (u \tilde\varphi k) 
\nonumber
\\
+ 2 (1+\sigma g_a)^2 \left( 
  (u p)_\| (u k)_\| + (u p)_\| (u \tilde\varphi k) + (u k)_\| (u \tilde\varphi p)
\right)
\Big]  , 
\\
\left. N^{\sigma \sigma'}_{\alpha \beta} \right|_{\sigma=-\sigma'}  
L_{\alpha \beta}^{\perp}
=
16 m_p m_n (1+\sigma) \Big[
(1+g_a^2) (u \tilde\Lambda u) - (1-g_a^2)
\Big]  
\Big[ 
(k \tilde\Lambda p) + (k \tilde\varphi p)
\Big]  .
\ee
In the selected frame of reference, where the nucleon medium is at rest 
[$ u=(1,0,0,0) $, $ (kp)_\| = (k \tilde\Lambda p) = k_0 p_0 - k_3 p_3 $], 
these expressions are 
\be
\left. N^{\sigma \sigma'}_{\alpha \beta} \right|_{\sigma=\sigma'}  
L_{\alpha \beta}^{\|}
=
16 m_p m_n (1 + \sigma g_a)^2 (p_0 + p_3) (k_0 + k_3)  , 
\\
\left. N^{\sigma \sigma'}_{\alpha \beta} \right|_{\sigma=-\sigma'}  
L_{\alpha \beta}^{\perp}
=
32 m_p m_n (1+\sigma) g_a^2 (p_0 + p_3) (k_0 - k_3)  . 
\ee
Hence, it is easy to obtain the final expression for the square of the 
amplitude (\ref{M}) of the process being calculated: 
\be
|M_\sigma|^2 
= 
4 (p_0 + p_3) \left[
(1 + \sigma g_a)^2 (k_0 + k_3) + 2 (1 + \sigma) g_a^2 (k_0 - k_3)
\right]  .
\ee
Note that in the case of nonrelativistic nucleons under consideration, the 
square of the amplitude depends only on the longitudinal electron and 
neutrino momentum components.

\section{
The neutrino absorption coefficient
}
\setcounter{equation}{0}

Below, we give the details of our calculation of the absorption 
coefficient ${\cal K}^{(\nu)}$ (\ref{K}) for process (\ref{1}) involving 
neutrinos. We used standard expressions for an elements of phase volume: 
$ dn^{(e)} = (dp_2^{(e)} dp_3^{(e)} L_y L_z) / (2 \pi)^2 $ for charged 
particles and $ dn = (d^3 p V) / (2 \pi)^3 $ for uncharged particles. The 
proton distribution function was assumed to be the Boltzmann one. 
Eliminating the integration over the proton momentum due to the 
$\delta^{(3)}$-function, we note that the absorption coefficient does not 
depend on the $p_2$ component of the electron momentum. Since 
$ p_2 = eB x_c $ in a magnetic field ($x_c$ is the centroid coordinate of 
the distribution of the electron wave function across the magnetic field), 
we can eliminate the integral over the $p_2$ component of the electron 
momentum: $ \int dp_2 = eB L_x $ The remaining $\delta$-function in energy 
can be simplified. Since the electrons and neutrinos are ultrarelativistic 
particles and the nucleons are nonrelativistic particles, their momenta 
are of the order of $ P^2 \sim T m $ (for nucleons) and $ p^2 \sim T^2 $ 
(for ultrarelativistic particles). Assuming the nucleon masses to be 
identical (where this does not cause any misunderstanding), we obtain 
\be
\delta \Big(
E_n - E_p^{(n)} + \omega - \varepsilon 
\Big) 
= \!\!\!\!\!\!\!
&&\delta \Big(
\frac {(\vec q)^2} {2 m_n} - \frac {( q_\| + k_\| - p_\| )^2} {2 m_p} 
- \frac{eBn} {m_p} + \omega - \varepsilon + ( m_n - m_p ) 
\Big)
\nonumber 
\\
&&\simeq 
\delta \Big(  
\frac {q_\perp^2} {2 m_n} - \frac{eBn} {m_p} 
+ \omega - \varepsilon + ( m_n - m_p ) 
\Big)  , 
\ee
where $q_\perp$ and $q_\|$ are the particle momentum components across and 
along the magnetic field, respectively. Since the dimensionless parameter 
$ b = (eB) / (2 m_p T) \ll 1 $ we may neglect this term in one of the 
$\delta$-function in the expression for the square of the $S$-matrix 
element (\ref{S}) by combining the two sums. In this case, the expression 
for the absorption coefficient is simplified and can be represented as 
\be
{\cal K}^{(\nu)} 
&=& 
\frac { 2 G_F^2 \cos^2 \theta_c } { (2 \pi)^4} eB 
\left( 
( 1 + 3  g_a^2 ) + ( 1 - g_a^2 ) k_\| / \omega 
\right) 
\int f_n F(q^2_\perp,b) d^3 q,  
\label{KK}
\\
F(q^2_\perp,b)  
&=& 
\int\limits_0^\infty \frac { dy } { 1 + \exp[\eta_e - y] } 
\sum_{n=0}^\infty \frac {1} {n!} \left( \frac{x}{b} \right)^n 
\exp \left[ - \frac {x} {b} \right] 
\nonumber
\\
&\times& 
\delta \left( 
x - bn - y + \frac { \omega + m_n - m_p } {T} 
\right) . 
\nonumber 
\ee
Here, $ x = q^2_\perp / ( 2 m_n T ) $ and $ y = \varepsilon / T $. Note 
that the function $F(q^2_\perp,b)$ has a simple asymptotic behavior for 
$ b \to 0 $, $bn$ is a finite number. The latter condition follows from 
the fact that as the magnetic field weakens, the number of Landau levels 
that contribute to the sum under consideration increases in inverse 
proportion to field strength. In this case, the sum over the proton Landau 
levels may be replaced by an integral. It can also be shown that in this 
limit, 
\be
\lim_{b \to 0 \atop bn \not \to 0}
\left[
\frac {1}  { n! b} \left( \frac{x}{b} \right)^n 
\exp \left[ - \frac {x} {b} \right]  
\right] 
=
\delta(x - bn) ,
\ee
where $\delta$ is the Dirac delta function. Hence, it is easy to find that 
\be
\lim_{b \to 0 \atop bn \not \to 0}
\left[
F(q^2_\perp,b) 
\right]
=
\frac {1} { 1 + \exp 
\bigg( 
\eta_e - (\omega + m_n - m_p)/T 
\bigg) } .
\ee
Since the function $F(q^2_\perp,b)$ does no depend on $q^2_\perp$ in the 
limit under consideration, the integration in (\ref{KK}) can be brought to 
the end. Given the definition of the neutron number density, 
$ N_n = 2/(2\pi)^3 (\int f_n d^3 q) $ we obtain the final expression for 
the neutrino absorption coefficient in the limit considered: 
\be
{\cal K}^{(\nu)} 
=
\frac { G_F^2 \cos^2 \theta_c } { 2 \pi } eB N_n 
\frac { ( 1 + 3  g_a^2 ) + ( 1 - g_a^2 ) k_\| / \omega } 
      { 1 + \exp \bigg( \eta_e - (\omega + m_n - m_p)/T \bigg)} .
\ee
The absorption coefficient for process (\ref{3}) involving antineutrinos 
can be calculated in a similar fashion to give 
\be
{\cal K}^{(\tilde\nu)} 
=
\frac { G_F^2 \cos^2 \theta_c } { 2 \pi } eB N_p 
\frac { ( 1 + 3  g_a^2 ) + ( 1 - g_a^2 ) k_\| / \omega } 
      { 1 + \exp \bigg( - \eta_e - (\omega - m_n + m_p)/T \bigg)} .
\ee

\section{
The force density along the magnetic field for the neutrino scattering by 
nucleons 
}
\setcounter{equation}{0}

In the low-energy limit, the effective Lagrangian for the neutrino 
scattering by nucleons is \cite{Raffelt} 
\be
{\cal L} = \frac {G_F} {\sqrt{2}} 
\Big(
\bar U_N ({\cal P'}) \gamma_\alpha (c_v + c_a \gamma_5 ) U_N ({\cal P})
\Big)
\Big(
\bar U_\nu (k') \gamma_\alpha (1 + \gamma_5 ) U_\nu (k)
\Big) .  
\ee
Here, $U_N$ and $U_\nu$ are the Dirac nucleon and neutrino bispinors; and 
$c_v$ and $c_a$ are the vector and axial constants of a neutral nucleon 
current. At low energies \cite{RafShekl}, 
$ c_v = - 1/2 $ and $c_a \simeq - 0.91/2 $ for neutrons and 
$ c_v = 0.07/2 $, $c_a \simeq 1.09/2 $ for protons. Using the 
nonrelativistic density matrix for the polarized nucleons, 
$ \rho = m_N ( 1 + \vec\xi \vec\sigma ) $, 
where $ \vec\xi = \pm \vec B / B $ is the nucleon polarization vector 
along the magnetic field and $\vec\sigma$ are the Pauli matrices, it is 
easy to obtain the following expression for the square of the $S$-matrix 
element for the neutrino scattering by nucleons per unit time: 
\be
\frac{|S_{if}|^2_\nu}{{\cal T}} 
= 
\frac { (2 \pi)^4 G_F^2 } { 2 V^3 \omega \omega' } 
\delta^{(4)} ( {\cal P} + k - {\cal P'} - k' ) 
\Big[
( c_v^2 + 3 c_a^2 ) \omega \omega' + ( c_v^2 - c_a^2 ) ( \vec k \vec k' ) 
\nonumber
\\
+ 2 c_v c_a ( S + S') ( \omega k'_\| + \omega' k_\| ) 
- 2 c_a^2 ( S - S' ) ( \omega k'_\| - \omega' k_\| ) 
\nonumber
\\
+ ( c_v^2 - c_a^2 ) S S' ( \omega \omega' + ( \vec k \vec k' ) ) 
+ 4 c_a^2 S S' k_\| k'_\|
\Big] .
\label{SN}
\ee
Here, $ k = (\omega, \vec k) $ and $ k' = (\omega', \vec k') $ are the 
4-momenta of the initial and final neutrinos; 
$k_\|$ and $k'_\|$ are the components of these momenta along the magnetic 
field; and $ S, S' = \pm 1 $ ($ S = ( \vec\xi \vec B ) / B $ are the 
components of the polarization vectors for the initial and final nucleons 
along the magnetic field). The square of the $S$-matrix element for the 
antineutrino scattering by nucleons can be derived from (\ref{SN}) by the 
change $ k \leftrightarrow k' $:  
\be
|S_{if}|^2_{\tilde\nu} = |S_{if}|^2_\nu ( k \leftrightarrow k' ) .
\ee 
Since we consider the neutrino scattering by nucleons in a strong magnetic 
field, for the subsequent calculations, it is necessary to take into 
account the contribution to the nucleon energy from the interaction of 
their magnetic moment with the magnetic field: 
$ E_N = m_N + {\cal P}^2 / (2 m_N) - (g_N S eB) / (2 m_N) $ where 
$g_N$ is the nucleon magnetic factor ($ g_n \simeq  -1.91 $ for the 
neutron and $ g_p \simeq 2.79 $ for the proton).

Next, we can simplify the $\delta$-function in energy. Using the fact that 
$ {\cal P}, {\cal P'} \sim \sqrt{m_N T} $ and $ k, k' \sim T $ and 
disregarding all terms $ \sim \sqrt{ T / m_N } $, we obtain 
\be
&&\delta \left( \omega - \omega' + E_N - E_N' \right) 
= 
\delta \left( 
\omega - \omega' + \frac { {\cal P}^2 } { 2 m_N } 
- g_N ( S - S' ) \frac {eB} { 2 m_N } 
- \frac { ( \vec {\cal P} + \vec k + \vec k' )^2 } { 2 m_N } 
\right) 
\nonumber
\\
&&\simeq 
\delta \left( 
\omega - \omega' - g_N ( S - S' ) \frac {eB} { 2 m_N } 
\right) .
\ee 
It is convenient to perform the subsequent calculations by separating out 
the contributions from $ S = S' $ and $ S = - S' $. Eliminating the 
$\delta$-function in momentum of the final nucleon and given that all 
terms linear in $k_\|$ and $k_\|'$ give no contribution to the momentum 
transfer to the medium along the magnetic field, we derive the following 
expression for the density of the force acting on a unit volume of the 
medium per unit time: 
\be
\Im _\| ^{(\nu)} 
&=&
\sum_S \left( 
\left. \Im _\| ^{(\nu)} \right|_{S=S'} 
+ \left. \Im _\| ^{(\nu)} \right|_{S=-S'}
\right) , 
\nonumber
\\
\left. \Im _\| ^{(\nu)} \right|_{S=S'} 
&=&
\frac { 2 G_F^2 c_v c_a } { ( 2 \pi )^8 } 
\int d^3 {\cal P} f_N ({\cal P}) 
\int d^3 k f_\nu (k) 
\int d^3 k' ( 1 - f_\nu (k') ) 
\nonumber
\\
&&\times 
\frac {S} { \omega \omega' }  
( \omega' k^2_\| - \omega k'^2_\| )  
\delta \left( \omega - \omega'\right) ,
\label{c5}
\\
\left. \Im _\| ^{(\nu)} \right|_{S=-S'} 
&=&
\frac { 2 G_F^2 c_a^2 } { ( 2 \pi )^8 } 
\int d^3 {\cal P} f_N ({\cal P}) 
\int d^3 k f_\nu (k) 
\int d^3 k' ( 1 - f_\nu (k') ) 
\nonumber
\\
&&\times 
\frac {S} { \omega \omega' }  
( \omega' k^2_\| + \omega k'^2_\| )  
\delta \left( \omega - \omega' - g_N S eB / m_N \right) .
\label{c6}
\ee 
Here, we use the fact that the nucleon gas is a Boltzmann one and, hence, 
$ 1 - f_N \simeq 1 $.

Since $ eB / ( m_N T ) \ll 1 $, the distribution function for the initial 
nucleons can be written as 
\be
f_N ({\cal P}) 
\simeq 
\bigg( 1 + g_N S eB / ( 2 m_N T ) \bigg) 
\exp \bigg[ - {\cal P}^2 / ( 2 m_N T ) + \eta_N \bigg] ,
\ee 
where $ \eta_N $ is the nonrelativistic nucleon degeneracy parameter. 
Interacting over the initial-nucleon momentum and using the fact that the 
nucleon number density is 
\be
N_N 
\simeq 
2 / (2 \pi)^3  
\int \exp \bigg[ - {\cal P}^2 / ( 2 m_N T ) + \eta_N \bigg] d^3 {\cal P} ,
\ee 
we obtain the following expression for (\ref{c5}) and (\ref{c6}) 
\be
\left. \Im _\| ^{(\nu)} \right|_{S=S'} 
= 
\frac { G_F^2 c_v c_a } { ( 2 \pi )^5 } 
N_N T^6 S \left( 
1 + g_N S \frac {eB} { 2 m_N T }
\right) 
I_1 , 
\\
I_1 
=
\int\limits^\infty_0 y^2 d y 
\int\limits^\infty_0 y'^2 d y'
\int\limits^1_{-1} d \chi 
\int\limits^{2 \pi}_0 d \varphi 
\int\limits^1_{-1} d \chi' 
\int\limits^{2 \pi}_0 d \varphi' 
f_\nu (y,\chi) \bigg( 1 - f_\nu (y',\chi') \bigg) 
\nonumber
\\
\times
\bigg( y' ( 1 - \chi^2 ) \cos^2 \varphi - y ( 1 - \chi'^2 ) \cos^2 \varphi' \bigg) 
\delta \left( y - y' \right) , 
\label{I1}
\\
\left. \Im _\| ^{(\nu)} \right|_{S=-S'} 
=
\frac { G_F^2 c_a^2 } { ( 2 \pi )^5 } 
N_N T^6 S \left( 
1 + g_N S \frac {eB} { 2 m_N T }
\right) 
I_2 , 
\\
I_2 
=
\int\limits^\infty_0 y^2 d y 
\int\limits^\infty_0 y'^2 d y'
\int\limits^1_{-1} d \chi 
\int\limits^{2 \pi}_0 d \varphi 
\int\limits^1_{-1} d \chi' 
\int\limits^{2 \pi}_0 d \varphi' 
f_\nu (y,\chi) \bigg( 1 - f_\nu (y',\chi') \bigg) 
\nonumber
\\
\times
\bigg( y' ( 1 - \chi^2 ) \cos^2 \varphi + y ( 1 - \chi'^2 ) \cos^2 \varphi' \bigg) 
\delta \left( y - y' - g_N S \frac {eB} { m_N T } \right) .
\label{I2}
\ee 
Here, $ y = \omega / T $, $ y' = \omega' / T $, $ \chi = \cos \theta $, 
and $ \chi' = \cos \theta' $, where $\theta$ and $\theta'$ are, 
respectively, the angles between the momenta of the initial and final 
neutrinos and the radial direction; and $\varphi$ and $\varphi'$ are the 
azimuthal angles. Since the square of the $S$-matrix element for the 
process involving antineutrinos can be derived by the change 
$ k \leftrightarrow k' $, it is easy to note that the expression for the 
momentum transferred along the magnetic field in the process of 
antineutrino scattering by nucleons is given by the formal change 
$ c_a^2 \to - c_a^2 $:  
\be
\Im _\| ^{(\tilde\nu)} 
= 
\Im _\| ^{(\nu)} ( c_a^2 \to - c_a^2 ) .
\ee 
For the integrals (\ref{I1}) and (\ref{I2}) to be calculated, we must use 
an explicit neutrino distribution function. We choose it in the form 
(\ref{fn}): $ f_\nu (k) = \Phi_\nu (r,\chi) F_\nu (\omega) $ where 
$ F_\nu (\omega) = ( 1 + \exp [ \omega / T - \eta_\nu ] )^{-1} $, 
$\eta_\nu$ is the neutrino degeneracy parameter.

It is convenient to express the final result in terms of the mean 
neutrino-flux parameters (\ref{J}), (\ref{chi}), and (\ref{e}). For the 
distribution function used, these parameters can be represented as 
\be
J_{\nu} 
&= &
(4 \pi)^{-1} \int \Phi_{\nu} (r,\chi) d \Omega 
\\
\langle \chi^2_\nu \rangle 
&=&
\left(
\int \chi^2 \Phi_\nu (r,\chi) d \Omega
\right)
\left(
\int \Phi_\nu (r,\chi) d \Omega
\right)^{-1}
\\
\langle \omega^n_\nu \rangle 
&=& 
\left(
\int \omega^{n+2} F_\nu (\omega) d^3 \omega 
\right)
\left(
\int \omega^2 F_\nu (\omega) d^3 \omega
\right)^{-1}
\ee

For completeness, we present the result of our calculation of the 
contributions to the force density from different nucleon polarizations 
separately: 
\be
&&\sum_S \left. \Im _\| ^{(\nu)} \right|_{S=S'} 
= 
- \frac{ G_F^2 g_N c_v c_a } { 2 \pi } 
\frac{ eB }{ m_N T } N_N N_\nu 
\langle \omega^3_\nu \rangle 
\bigg( \langle \chi^2_\nu \rangle - 1 / 3 \bigg) ,
\\
&&\sum_S \left. \Im _\| ^{(\nu)} \right|_{S=-S'} 
= 
- \frac{ G_F^2 g_N c^2_a } { 2 \pi } 
\frac{ eB }{ m_N T } N_N N_\nu 
\Bigg[
T \langle \omega^2_\nu \rangle 
\bigg( \langle \chi^2_\nu \rangle - 1 / 3 \bigg) -
\nonumber 
\\
&&- \bigg(
\langle \omega^3_\nu \rangle - 5 T \langle \omega^2_\nu \rangle
\bigg)
\bigg(
5 / 3 -  \langle \chi^2_\nu \rangle
\bigg)
+ 2 J_{\nu}
\bigg(
\langle \omega^3_\nu \rangle -
5T_\nu \langle \omega^2_\nu \rangle
\bigg)
\bigg(
1 - \langle \chi^2_\nu \rangle
\bigg)
\Bigg] .
\ee 
It is easy to verify that each of these expressions is zero for a thermal 
equilibrium of the neutrinos with the medium ($ T_\nu = T $, $ J_\nu = 1 $, 
$\langle \chi^2_\nu \rangle = 1 / 3 $), as follows from fundamental 
physical principles.


\end{document}